\documentclass[english,preprint]{revtex4}
\usepackage[latin1]{inputenc}
\usepackage{graphicx}
\pdfoutput=1
\usepackage{amsmath, amssymb,empheq,xcolor}
\usepackage[T1]{fontenc}
\usepackage[small,]{caption}
\usepackage{hyperref}
\makeatletter
\usepackage{babel}
\makeatother

\begin{document}

\title{Nano-lens diffraction around a single heated nano particle}
\author{Markus Selmke, Marco Braun, Frank Cichos}
\address{Molecular Nanophotonics Group, Institute of Experimental Physics I, University Leipzig, 04103 Leipzig, Germany}

\date{\today}

\begin{abstract}
The action of a nanoscopic spherically symmetric refractive index profile on a focused Gaussian beam may easily be envisaged as the action of a phase-modifying element, i.e.\ a lens: Rays traversing the inhomogeneous refractive index field $n\left(\mathbf{r}\right)$ collect an additional phase along their trajectory which advances or retards their phase with respect to the unperturbed ray. This lens-like action has long been understood as being the mechanism behind the signal of thin sample photothermal absorption measurements \cite{Jurgensen1995,Moreau2006}, where a cylindrical symmetry and a different lengthscale is present. In photothermal single (nano-)particle microscopy, however, a complicated, though prediction-wise limited, electrodynamic (EM) scattering treatment was established \cite{Berciaud2006} during the emergence of this new technique. Our recent study extended \cite{2011arXiv1105.3815S} this EM-approach into a full ab-initio model describing the reality of the situation encountered and showed for the first time that the mechanism behind the signal, despite its nanoscopic origin, is also the lens-like action of the induced refractive index profile only hidden in the complicated guise of the theoretical Mie-like framework. The diffraction model proposed here yields succinct analytical expressions for the axial PT signal shape and magnitude and its angular distribution, all showing the clear lens-signature. It is further demonstrated, that the Gouy-phase of a Gaussian beam does not contribute to the relative photothermal signal in forward direction, a fact which is not easily evident from the more rigorous EM treatment. The model may thus be used to estimate the signal shape and magnitude in photothermal single particle microscopy.
\end{abstract}
\maketitle
\section{Introduction}
Photothermal lens spectroscopy (PLS) has become a valuable tool in the study of solids and liquids \cite{Lu2010LabelFree,Sinha2000,Brusnichkin2007}. Recent publications include the study of non-linear effects \cite{Battaglin2001} and nanoparticles in solution \cite{Brusnichkin2007}. Many authors have focused on the theoretical description of the thin-sample slab geometry which is often utilized in such macroscopic lensing experiments, often providing numerical equations which may be used to obtain absorption coefficients \cite{Jurgensen1995,Moreau2006}. In all these models the thermal lens induced originates from the absorbed power of a heating laser which constitutes a spatially extended cylindrically symmetric heat source in the heat equation. The solution obtained is then used to study the effect on the propagation of a probing laser beam. In thermal lens spectroscopy (TL) the probing beam is coaxial (possibly offset) with the heating beam \cite{Moreau2004, Moreau2006} while in beam deflection spectroscopy \cite{Harada1993,Wu1991} (BDS) the probing and the heating beams are aligned perpendicularly to each other. Both methods have in principle the same sensitivity \cite{Jackson1981}.

In contrast to these macroscopic techniques, a new microscopic approach has been developed in the 1990's by Harada and Kitamori \cite{Harada1993AnalChem, Kitamori2000} termed photothermal lens microscopy. The developement of single particle photothermal microscopy \cite{Orrit2010Science,Lounis2004,Boyer2002} followed and detects a very different kind of thermal lens with a mildly modified standard confocal fluorescence microscope: Instead of an axially symmetric refractive index profile a spherically symmetric profile $n\left(r\right)$, created by the point-like heat-source of the absorbing nano-particle, is probed. Also, instead of a profile that decays on the length-scale of the heating beam focus, in PT single particle microscopy a lens is probed which decays to half its value on the length-scale of the nanoparticle. On the other hand, the profile extends infinitely as $1/r$ and thus misses a characteristic length scale. This is the reason, why the description presented in this paper and within the recent EM-study \cite{2011arXiv1105.3815S} provide no evidence for a role-play of the Gouy-phase which is otherwise important for the probing of small scatterers as shown by Hwang and Moerner \cite{Hwang2007}.
While models for spherical absorbers have been put forward \cite{Andika2010,Harada1994}, these were numerical in nature and targeted for large $\mu\rm m$-sized absorbers. The theoretical description of the nanoscopic photothermal lens has been first given by Berciaud et al. \cite{Berciaud2006} in a scattering treatment relying on an extinction mechanism. Our recent ab-initio theoretical description of the electrodynamic problem has shown, however, that instead a simple lensing mechanism is responsible for the photothermal signal of nanoscopic absorbers showing a clear angular thermal diffraction signature and a double-lobe lens signature in axial scans while disproving the assumption of interference-dominance in forward detection in the general situation.

It is thus the aim of this paper to show that both worlds, the vast literature on macroscopic thermal lens spectroscopy and the recent emerging tool of single particle thermal lens microscopy are very similar. The diffraction picture which has been a successful tool in the first domain will be shown to yield analytical and easily tractable expressions in the nano-scopic domain. They will allow for a quantitative assesment of absorption cross-sections of single nano-object based on standard photothermal measurements and will be able to explain the main phenomena of photothermal microscopy qualitatively as well as quantitatively, while providing an intuitive picture of the working mechanism. The quality of the simple model is checked against the more elaborate electromagnetic model within the extended scattering description. Axial scans and angular patterns of single heated nano particles will be described and compared. The angular diffraction pattern will be shown to explain the signal inversion observed for the first time for a single nano-particle upon the introduction of an inverse aperture in the detection path.

\section{The lens}
The lens to be considered in single particle generated nano-lens experiments such as photothermal microscopy originates from the absorption of optical power provided by a focused laser beam by a single nano-particle. The absorbing particle can be treated as a point-like heat-source, yielding the steady-state temperature profile $T\!\left(\mathbf{r}\right)=T_0 + \Delta T_0 R/r$ which decays with the inverse distance $r$ from the particle of radius $R$. In the case of modulated heating, as utilized in the lock-in approach common to photothermal single particle microscopy, the assumption remains valid as long as the modulation frequencies used remain below $1\,\rm MHz$, typically \cite{Berciaud2006, 2011arXiv1105.3815S}. By the temperature dependence of the surrounding mediums' refractive index $n$, a corresponding refractive index profile $n\left(\mathbf{r}\right)$ is established:

\begin{equation}
n\left(\mathbf{r}\right)=n_0+\frac{\mathrm{d}n}{\mathrm{d}T} \Delta T\left(\mathbf{r}\right)= n_0 +\Delta n \frac{R}{r}, \label{EqRefracProfile}
\end{equation}

with $\Delta n= \Delta T_0 \left(\mathrm{d}n/\mathrm{d}T\right)$ being the heating induced refractive index contrast and $n_0=n\left(T_0\right)$ the unperturbed refractive index. The amount of energy absorbed by the particle is determined by its absorption cross-section $\sigma_{\rm abs}$ and the intensity of the heating laser $I_{h}\left(z_p\right)$ at the particle position. Together with the thermal conductivity $\kappa$ and the radius $R$ of the particle this controls the induced temperature and thus the contrast $\Delta n$ of the lens

\begin{equation}
\Delta n = \frac{\sigma_{\rm abs}\,I_{h}}{4\pi \kappa R}\left(\frac{\mathrm{d}n}{\mathrm{d}T}\right)\label{EqRefrcativeContrast}.
\end{equation}

The lens described by Eqn.\ \ref{EqRefracProfile} decays to half its maximum perturbation $\Delta n$ at a distance of $r=2R$, making it a nanoscopic object. Nonetheless, it has infinite extent. It will only be limited by the probing beam which will for most cases be a confined and focused beam of diffraction limited extent. As $\Delta n$ will quantify the photothermal single particle signal, the knowledge of the heating beam intensity and the thermal properties of the embedding bulk material will then allow the determination of the particles' absorption cross-section $\sigma_{\rm abs}$. It is thus necessary to obtain expressions for the photothermal single particle signal from the induced refractive contrast $\Delta n$. This is the purpose of the next section.

\section{The diffraction integral\label{diffraction}}

\begin{figure}[h]
  \centering
 	\includegraphics[scale=0.7]{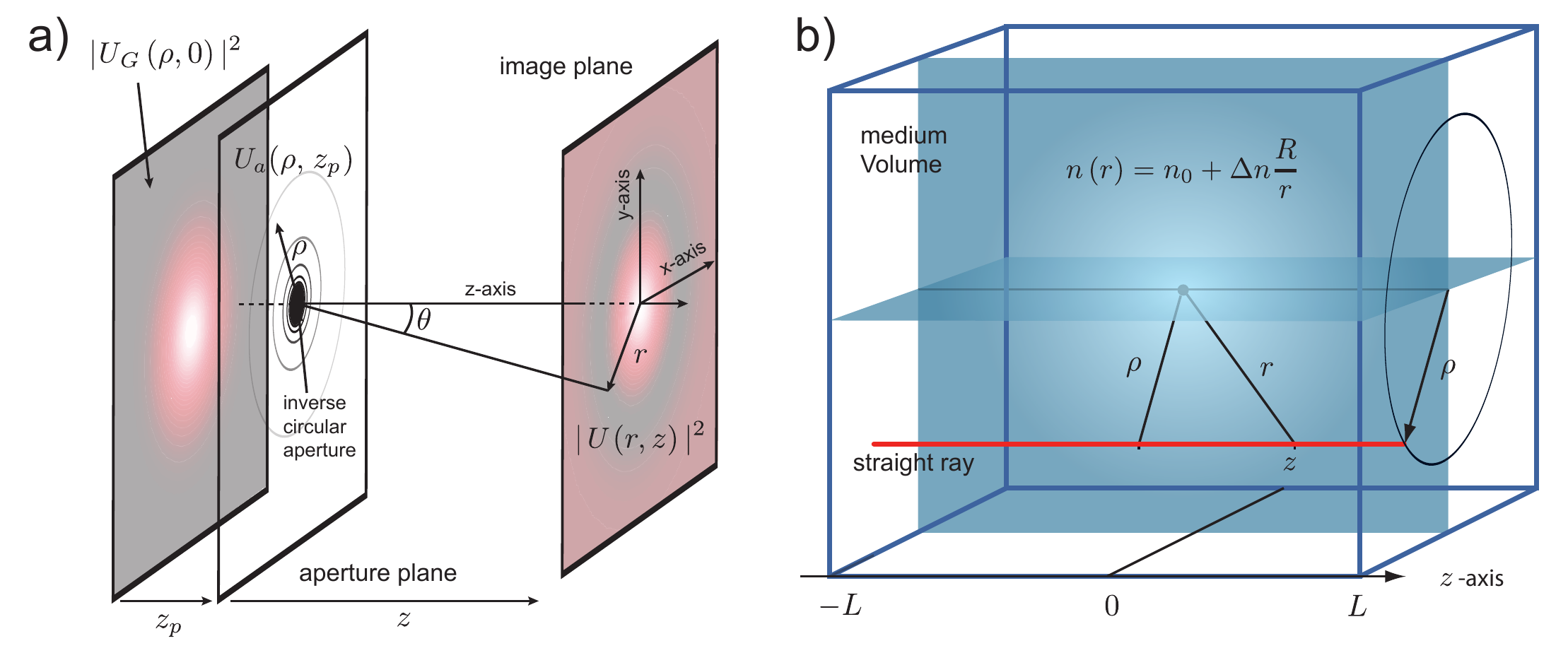}
  \caption{The scalar diffraction model. a) geometry for the diffraction integral (Eq. \ref{EqDiffractionIntegral}). The shading in the image-plane corresponds to the difference in intensities between the diffraction of a cold and a hot nano-particle, Eq.\ \ref{EqSignalOffAxis}. b) geometry for the computation of the phase advance (Eq. \ref{EqPhaseAdvance})}
  \label{fig:fig1}
\end{figure}

The Frauenhofer diffraction integral for circlular apertures in the Fresnel-grade approximation connects the complex field-amplitude of the probe beam $U_a\left(\rho\right)$ in the aperture plane at $z=0$ to the complex field amplitude $U\left(r,z\right)$ in the image plane at a distance $z$ (see Figure \ref{fig:fig1}, a)) \cite{Horvath2003,Teng2007}. A further factor is included which represents the collected phase $\Delta \Phi$ for each component-wave as a result of the thermal lens (see Figure \ref{fig:fig1}, b)) \cite{Vigasin1977}.

\begin{equation}
U\left(r,z\right)=\frac{k}{i z}\exp\left(-i \frac{k r^2}{2 z}-i k z\right)\int_{R}^{\infty}\!\!\!\!U_a\left(\rho\right)\, \exp\left(-i \frac{k \rho^2}{2 z}\right) J_0\left(\frac{k \rho r}{z}\right) \exp\left(-i\Delta \Phi \left(\rho\right)\right)\rho \mathrm{d}\rho\label{EqDiffractionIntegral}
\end{equation}

To simplify matters, we will consider an axially symmetric scenario only, i.e. the refractive index profile will be symmetric with respect to the optical axis. This corresponds to the case of a heated nano-particle positioned along the optical axis (see Figure\  \ref{fig:fig1}, a)).

The phase advance $\Delta \Phi \left(\rho\right)$ for a single ray passing the lens immersed in a sample slab of thickness $L$ at a distance $\rho$ from the optical axis can be approximated via a straight ray calculation already utilized for collimated beams in \cite{Vigasin1977}:
\begin{equation}
\Delta \Phi \left(\rho\right) =k_0\int_{-L}^{L}\left[n_0 + \frac{R\,\Delta n}{\sqrt{z^2+ \rho^2}}\right] \mathrm{d} z \approx 2k_0\left[L n_0 - R\,\Delta n \ln\left(\frac{\rho}{2L}\right)\right]\label{EqPhaseAdvance},
\end{equation}
where $\sinh^{-1}(z)=\log\left(z+\sqrt{1+z^2}\right)$ and $L\gg \rho$ were used after the integration. Although the integration in the distance $\rho$ extends to infinity, the weighting by the Gaussian field amplitude $U_a\left(\rho\right)$ will ensure the validity of the inequality.

The field amplitude in the aperture plane will be taken to be the complex field of the probing Gaussian beam \cite{SalehTeich} having a focus displaced by $z_p$ from the center of the heated particle (see Figure \ref{fig:fig1}, a)):
\begin{equation}
U_a\left(\rho\right)=U_0\frac{\omega_0}{\omega\left(z_p\right)}\exp\left(-\frac{\rho^2}{\omega^2\left(z_p\right)}\right)\exp\left(-ik z_p - i \frac{k \rho^2}{2R_C\left(z_p\right)}+i \zeta_G\left(z_p\right)\right),
\end{equation}

where the beam-waist in the aperture plane is $\omega\left(z_p\right)=\omega_0 \left[1+z_p^2/z_R^2\right]^{1/2}$, the curvature is given by $R_C\left(z_p\right)=z_p\left[1+z_R^2/z_p^2\right]$ and the Gouy-phase $\zeta_G\left(z_p\right) =\tan^{-1}\left(z_p/z_R\right)$. The Rayleigh-range $z_R$ is connected to the beam-waist $\omega_0$ and the wave-number $k=k_0 n_0$ by $z_R=k \omega_0^2/2$, where $k_0=2\pi/\lambda$ represents the vacuum wave-vector for the wavelength $\lambda$.

To put photothermal single particle microscopy on a quantitative footing \cite{2011arXiv1105.3815S}, we have introduced the relative PT signal $S$. It is the change in intensity $I\propto \left|U\right|^2$ within the image-plane relative to the constant much larger background of the unperturbed field:

\begin{equation}
S\left(r,z\right)=\left[\left|U\left(r,z\right)\right|_{\Delta n\left(\Delta T_0\left(z_p\right)\right)}^2-\left|U\left(r,z\right)\right|_{\Delta n=0}^2\right] \,\Big{/}\, \left|U\left(r=0,z\right)\right|_{\Delta n=0}^2\label{Eq:Signal}
\end{equation}

This relative signal will be independent of the probe power $P_d$ as long as no additional heating is induced, meaning that $\Delta n=\Delta n\left(P_h\right)=\rm const.$ with $P_d$. In case of a single laser being diffracted by its own induced thermal lens, the relative signal will be proportional to $P_d$.

\begin{figure}[h]
  \centering
 	\includegraphics[scale=0.75]{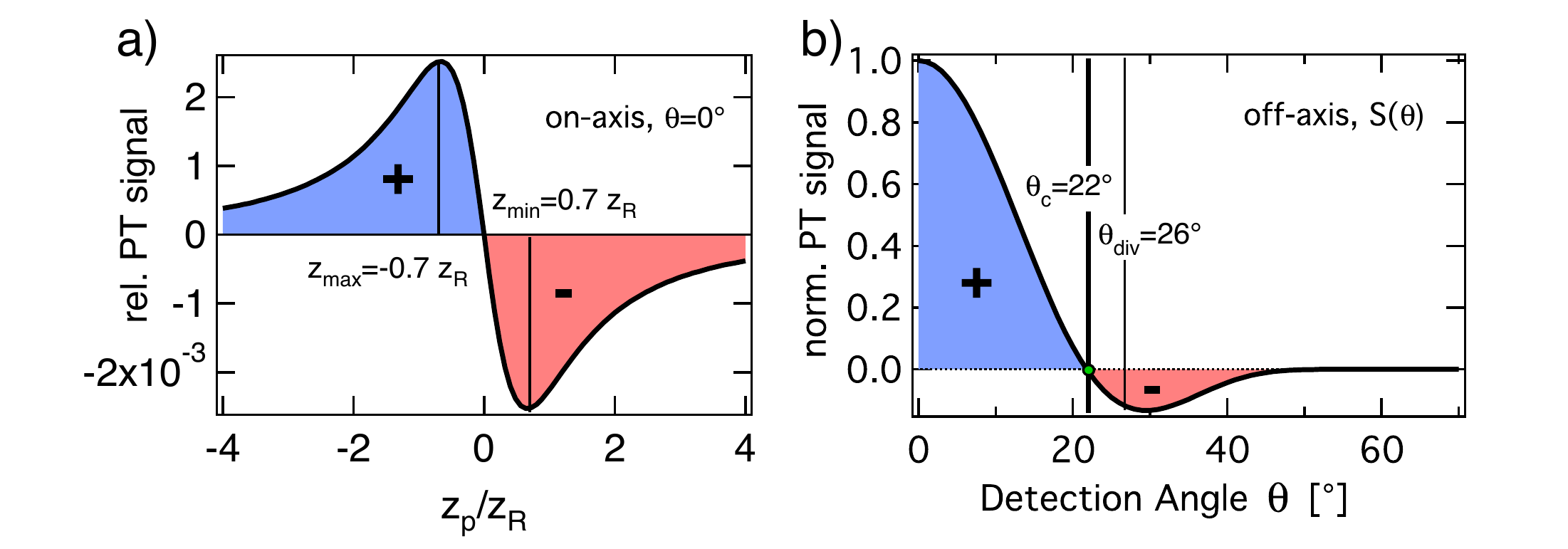}
  \caption{a) z-scan of the rel. photothermal signal for on-axis detection, i.e. $\theta = 0^\circ$ , Eqn.\ \ref{EqSignalOnAxis}. b) Angular pattern of the photothermal signal for $z_p=-z_R$, i.e. Eqn.\ \ref{EqSignalOffAxis}. The plot has been normalized to 1.0 on the optical axis. The parameters used in a) and b) the same as detailed in Figure\ \ref{fig:fig3Compare}.}
  \label{fig:fig1zScan}
\end{figure}

Through direct computation it can be shown that for small particles $R<100\,\rm nm$ the relative signal is not affected much by the geometrical diffraction of the inverse aperture disc of radius $R$. The diffraction-integral may be extended to $R=0$ without an appreciable change in the result. We find the following analytic expression on the optical axis ($r=0$), corresponding to the intensity-change detected in a closed aperture scenario:

\begin{equation}
S\left(z_p\right)=\left|\zeta^{-i R \Delta n k_0} \Gamma\left(1+i R \Delta n k_0\right)\right|^2-1,\label{EqSignalOnAxis}
\end{equation}

with the abbreviation $\zeta\left(z,z_p\right)=\left(1/\omega^2\left(z_p\right)+i k/\left(2 z\right)+i k/\left(2R_C\!\left(z_p\right)\right)\right)$ and $\Gamma\left(z\right)$ being the complex valued gamma function. The units of this expression seem to be odd at first, but a factor of the units meter $[\rm m]$ to the appropriate power has been omitted and ensures the unit less-ness of the expression. This expression being valid in the intermediate zone, the dependence on the image-plane distance $z$ may be dropped in the far-field region by leaving out the second summand of $\zeta$. The resulting dispersion-like signal shape (see Figure\ \ref{fig:fig1zScan} a)) obtained through this expression may have been anticipated from the close similarity of the situation presented here and the thermal lens model used in thin sample slab geometries \cite{HuWhinnery1973,Moreau2006,Jurgensen1995}. In these cases, the parabolic refractive index profile approximation for example yields an optimal probing beam offset relative to the sample of one confocal distance, $z_p=z_R$. Here, the maximum signal is obtained when the heated particle is offset by about $z_p=0.7 z_R$.
 For the off-axis image-plane signal, where the particle is still assumed to be positions on the optical axis, the radial coordinate $r$ in the image-plane was transformed to an angular coordinate via $\tan\left(\theta\right)=r/z$. The expression for the signal (Eqn.\ \ref{Eq:Signal}) detected under an angle $\theta$ for $r\ne 0$ then reads:

\begin{eqnarray}
S\left(\theta,z_p\right)&=&\exp\left(\frac{-k^2 \tan^2 \theta\,\textnormal{Re}\left(\zeta^{-1}\right)}{2}\right)\times\nonumber\\
&&\left[\left|\zeta^{-i R \Delta n k_0}\Gamma\left(1+iR \Delta n k_0\right) {}_1 F_{\,1}\left(-i R \Delta n k_0,1,\frac{k^2 \tan^2 \theta}{4 \zeta}\right)\right|^2-1\right].\label{EqSignalOffAxis}
\end{eqnarray}

${}_1 F_{\,1}$ denotes the confluent hypergeometric function of the first kind. It may be related to the complex-order Laguerre polynomial via ${}_1 F_{\,1}\left(-a i,1,b\right)=L_{i a}\left(b\right)$ for $a\in$ Reals. For small detection angles $\theta \rightarrow 0$ the expression reduces to the on-axis expression Eqn.\ \ref{EqSignalOnAxis} since $\tan\left(\theta\right)\rightarrow 0$ and ${}_1 F_{\,1}\left(c_1,1,0\right)=1$ for any number $c_1$. For a probing beam positioned behind the lens ($z_p <0$), the angular pattern described by this equation shows a peak towards the center and an annular dip at larger angles ($\theta_c=22^\circ$, see Figure\ \ref{fig:fig1zScan} b)). For the case of a probing beam being positioned in front of the lens ($z_p >0$), the angular pattern changes its sign, relative to the previous scenario. Overall, the energy of the beam is only redistributed by the action of the lens, i.e. an integration of $S\left(\theta,z_p\right)\sin\left(\theta\right)$ from $0$ to $\pi/2$ gives zero.

From this result we may, similar to the treatment given in \cite{Escalona2008}, compute the total detected photothermal signal if a finite detection aperture is used. In this case it is not the modulation of the intensity detected on-axis or under a certain angle $\theta$, but rather one needs to integrate the angular spectrum given in Eqn.\ \ref{EqSignalOffAxis} over the angular detection domain:

\begin{equation}
S\left(\theta_{\rm min},\theta_{\rm max},z_p\right)=F\times 2\pi\int_{\theta_{\rm min}}^{\theta_{\rm max}}S\left(\theta,z_p\right)\sin\left(\theta\right)\cos^{-3}\left(\theta\right)\mathrm{d}\theta ,\label{EqSignalNAd}
\end{equation}

wherein the factor $F$ compensates for taking the constant background $\left|U\left(r=0,z\right)\right|_{\Delta n=0}^2$ in the integrand for the normalization in Eqn.\ \ref{EqSignalOffAxis} instead of the true Gaussian intensity.
\begin{equation}
F=\frac{1}{A_{\rm sphc}}\frac{P_{d,I_0}}{P_{d,I}}=\frac{2\,z_R^2}{\pi\omega_0^2}\left[\exp\left(-2\tan^2\left(\theta_{\rm min}\right)z_R^2/\omega_0^2\right)-\exp\left(-2\tan^2\left(\theta_{\rm max}\right)z_R^2/\omega_0^2\right)\right]^{-1} \label{EqSignalNAd}
\end{equation}

A change of integration variables has been done from $r$ to $\theta$ via $\tan\left(\theta\right)=r/z$ and the power $P_r$ contained within a radius $r$ at a distance $z$, which is taken large as compared to the Rayleigh-range $z_R$, was used as given by $P_r\left(z\right)=P_0\left[1-\exp\left(-2r^2/\omega^2\left(z\right)\right)\right]$. Further, the expressions $P_0=\pi I_0 \omega_0^2/2$ and $I\left(z\right)=I_0 \omega_0^2/\omega^2\left(z\right)$ were used for the fraction of the power collected in the spherical cap area $A_{\rm spec}$ as obtained by taking a constant on-axis intensity, $P_{d,I_0}=A_{\rm spec} I\left(z\right)$, and by using the correct varying intensity as determined by the Gaussian beam. Thus, if only the central bump of the diffraction pattern is collected, as it is usually the case when a dry objective collects the high-NA focused probe beam, a z-scan will show a clear change in sign. Indeed, a z-scan as described by Eqn.\ \ref{EqSignalOnAxis} and Eqn.\ \ref{EqSignalNAd} shows a change in sign and a dispersion-like signal if $\Delta n$ is changed according to the axial shifting of the absorbing particle with respect to the local heating laser. This is just what one would expect for the collected signal behind a (modifiable) diverging lens if probed by a focused beam and the lens were moved along the beam axis across the focus while its focal length were to change in such a way that it has its smallest $|f|$ close to the probe beam focus and gets weaker, i.e.\ $|f|$ larger, for large offsets.
A further consequence of the given angular pattern of the photothermal signal is the possibility of an inversion of the photothermal signal. By using an annular aperture, only the annular region described by Eqn.\ \ref{EqSignalOffAxis} may be collected, and the total collected signal as described by Eqn.\ \ref{EqSignalNAd} will be of opposite sign as compared to the use of a usual circular aperture (Objective NA). Indeed, the reversal of sign in the photothermal signal could both be observed experimentally for $R=60\,\rm nm$ AuNPs and confirmed in the calculations using the GLMT (see section \ref{ChapterGLMT}).

Equations (\ref{EqSignalOnAxis}) and (\ref{EqSignalOffAxis}, \ref{EqSignalNAd}) present the main results of this paper and give the background normalized intensity change $S=\Delta I/I$ detected either on the optical axis or detected under a finite angle $\theta$ with respect to the optical axis upon the introduction of the lens $n\left(\mathbf{r}\right)$ (Eqn. \ref{EqRefracProfile}). Hereby, the particle/lens was assumed to be displaced by a distance $z_p$ with respect to the probing beam-waist position (see Figure\ \ref{fig:fig1}). It seems worth mentioning that in this approach the Gouy-phase terms, while present in the fields in the aperture plane, cancel each other in the relative photothermal signal since they do not depend on the integration variable $\rho$ and may thus be taken in front of the integrals.
The found expressions were used to generate the plots in Figures \ref{fig:fig1zScan} and \ref{fig:fig3Compare} and allow the study of z-scans in photothermal microscopy setups similar to the thin sample slab studies in \cite{Escalona2008,Mian2002zscan,Gnoli2005,Gnoli2007,Battaglin2001,Polloni2003}. To this end, the collection angle depends on the collecting microscope objectives' numerical aperture through $\theta_{\rm max}=\arcsin\left({\rm NA_d}/n_m\right)$, while the illuminating objective determines the beam waist(s).

\section{Comparison to rigorous Vectorial EM treatment\label{ChapterGLMT}}

\begin{figure}[h]
  \centering
 	\includegraphics[scale=0.9]{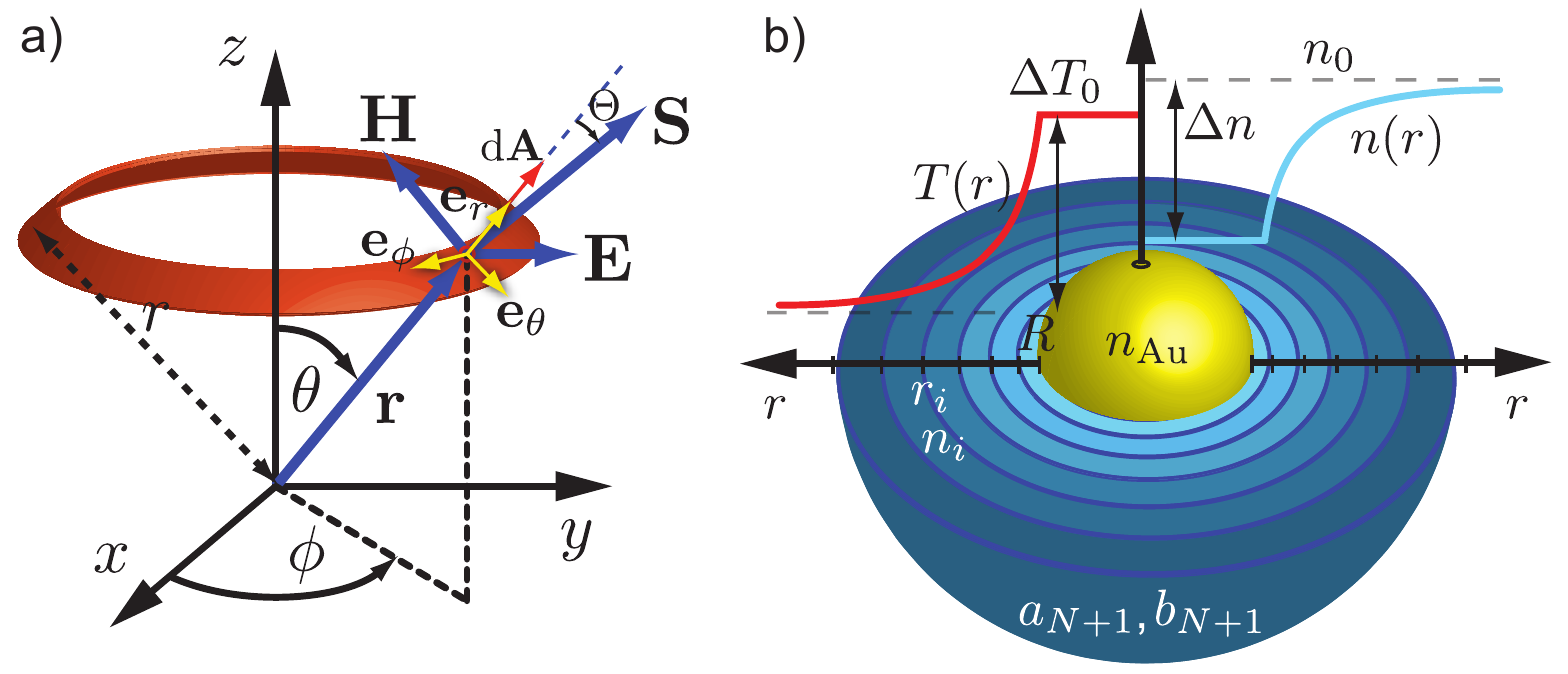}
  \caption{a) Schematic of the Poynting-vector integration within the GLMT framework. b) Discretization of the refractive index profile $n\left(r\right)$ into concentric spherical shells for the computation of the multilayer Mie scatter coefficients \cite{Pena2009}.}
  \label{fig:fig2}
\end{figure}

To compare the found results Eqn. (\ref{EqSignalOnAxis} and \ref{EqSignalOffAxis}) to a more rigorous solution of the problem, a full vectorial electromagnetic treatment will be used in the following. Therefore, the problem at hand may be expressed as the scattering of a shaped incident probe field interacting with a multilayered scatterer. The scatterer described by Eqn.\ (\ref{EqRefracProfile}) may be viewed as an unbounded gradient refractive index lens (GRIN). The scattering of such an object has been studied in the literature and a publicly available C-code is attainable through reference \cite{Pena2009} providing the scattering coefficients of the GRIN $a_{N+1}$ and $b_{N+1}$ when the scatterer discretized into $N$ concentric spherical shells (Figure \ref{fig:fig2}, b)). The generalized Lorenz-Mie theory (GLMT) which is applicable to any spherically symmetric scatterer \cite{Gousebet1995GLMTMLS} positioned in an arbitrarily shaped beam \cite{Gouesbet1995,Gouesbet1988,Gouesbet2011}. We have therefore adopted and modified it to yield an analytical expression for the power $P_d$ of the total electromagnetic field ($\mathbf{E}^t$, $\mathbf{H}^t$) detected in the far-field under a scattering-angle $\theta$. This treatment has been verified by exhaustive comparison to single gold nanoparticle scattering and photothermal microscopy \cite{2011arXiv1105.3815S}. The quantity of interest is the projected Poynting-vector $\mathbf {S}^{t}$ under a certain angle $\theta$ with respect to the optical axis (Figure \ref{fig:fig2}, a)):

\begin{equation}
\mathrm{d}P_d\left(\theta,\phi\right)=\mathbf{S}^{t}\cdot \mathrm{d}\mathbf{A}=S^{t}_{\perp}\mathrm{d}A\label{eq:PdbyPoynting}.
\end{equation}

Within the GLMT formalism the total field is mathematically decomposed into series expressions of the incidence field $\mathbf{E}^i$ and $\mathbf{H}^i$ with complex expansion coefficients $g_n$, and an outgoing scattered field $\mathbf{E}^s$ and $\mathbf{H}^s$ described by complex scatter coefficients $a_n$ and $b_n$. Thereby, the total field consists of the incident field and the scattered field, $\mathbf{V}^t=\mathbf{V}^i+\mathbf{V}^s$ for $\mathbf{V}=\mathbf{E}$ or $\mathbf{H}$. The scattered far-field is commonly expressed via the scattering amplitudes $S_{1,2}$ through $E_\theta^s=\frac{i E_0}{kr}\exp\left(-ikr\right)\cos\left(\phi\right)S_{2}\left(\theta\right)$, $H_\theta^s=-\frac{H_0}{E_0}E_{\phi}^s$, $E_\phi^s=\frac{-iE_0}{kr}\exp\left(-ikr\right)\sin\left(\phi\right)S_{1}\left(\theta\right)$, $H_\phi^s=\frac{H_0}{E_0}E_{\theta}^s$. These amplitudes are:

\begin{equation}
S_{1 \atop 2}=\sum_{n=1}^{\infty}\frac{2n+1}{n\left(n+1\right)} g_n\left[a_n {\Pi_n\left(\cos\theta\right) \atop  \tau_n\left(\cos\theta\right)}+b_n {\tau_n\left(\cos\theta\right) \atop  \Pi_n\left(\cos\theta\right)}\right],
\end{equation}

wherein the angular functions are $\Pi_n^m\left(\cos\theta\right)=P_n^m\left(\cos\theta\right)/\sin\theta$, $\tau_n^m\left(\cos\theta\right)=\mathrm{d}P_n^m\left(\cos\theta\right)/\mathrm{d}\theta$ and $P_m^m$ are the associated Legendre polynomials.
This artificially decomposes the detected power into three parts which are not individually detectable in the scattering situation:  $\mathrm{d}P_d=\mathrm{d}P_{\rm inc}+\mathrm{d}P_{\rm sca}+\mathrm{d}P_{\rm ext}$. Now, the time-averaged projected Poynting vector may be computed via $\langle S^{t}_{\perp}\rangle=\frac{1}{2}Re(E_\theta^t H_\phi^{t*}-E_\phi^t H_\theta^{t*})$, yielding three terms:

\begin{equation}
2\langle S^{t}_{\perp}\rangle=Re\left(E_\theta^i H_\phi^{i*}-E_\phi^i H_\theta^{i*}\right)+Re\left(E_\theta^s H_\phi^{s*}-E_\phi^s H_\theta^{s*}\right)+Re\left(E_\theta^i H_\phi^{s*}+E_\theta^s H_\phi^{i*}-E_\phi^i H_\theta^{s*}-E_\phi^s H_\theta^{i*} \right).
\end{equation}

The shape of the incidence field is determined by complex-valued expansion coefficients $g_n$, the so-called beam shape coefficients (BSCs). These coefficients reduce to $g_n=1$ for plane-wave illumination, i.e. regular Mie Theory. For a particle illuminated on-axis by a weakly focused Gaussian beam the modified local approximation (MLA, \cite{Gousebet1995modlocApprox}) has been developed:
\begin{equation}g_n\left(z_p\right)=Q\exp\left(-Q\,s^2\left(n-1\right)\left(n+2\right)\right)\exp\left(i \gamma s^{-1}/2\right).\label{eqngn}\end{equation} 

Herein, $Q=\left(1+i s \gamma\right)^{-1}$ with the beam-confinement factor $s$ defined through $s=\omega_0/(2z_R)$ and the defocussing parameter $\gamma=2 z_p/\omega_0$ describes the displacement of the particle relative to the beam-waist $\omega_0$, with $z_p <0$ being the situation where the focus is between the particle and the collecting objective. The result of Equation (\ref{eq:PdbyPoynting}), with the integration over the azimuthal angle $\phi$ carried out, in the far-field can now be expressed as differential cross-sections in any forward/backward polar angle $\theta$, $\mathrm{d}P_{\rm ext}\left(\theta\right)=-\mathrm{d}\sigma_{\rm ext}\left(\theta\right) I_0$ and $\mathrm{d}P_{\rm sca}\left(\theta\right)=\mathrm{d}\sigma_{\rm sca}\left(\theta\right) I_0$, where the Gaussian beam focus intensity $I_0=2P_0/\pi\omega_0^2$ was used:

\begin{equation}
\mathrm{d}\sigma_{\rm sca}\left(\theta\right)=\, \frac{\pi}{k^2}\left(|S_{1}\left(\theta\right)|^2+|S_{2}\left(\theta\right)|^2\right), \quad \mathrm{d}\sigma_{\rm ext}\left(\theta\right)=\frac{\pi}{k^2}\left[\textnormal{Re}\left(M\right)\,\textnormal{Re}\left(S_{12}\right)+\textnormal{Im}\left(M\right)\,\textnormal{Im} \left(S_{12}\right)\right].\label{eqnISI}
\end{equation}

The auxilliary functions $S_{12}=S_1\left(\theta\right)+S_2\left(\theta\right)$ and $M$ were introduced:
\begin{equation}
M\left(\theta\right)=\sum_{n=1}^{\infty}\,\frac{2n+1}{n\left(n+1\right)}g_n\, \left[\Pi_n\left(\cos\theta\right)+\tau_n\left(\cos\theta\right)\right].
\end{equation}

To obtain the total cross-sections, one needs to compute $\sigma_{\rm sca}=\int_{0}^{\pi}\mathrm{d}\sigma_{\rm sca}\left(\theta\right)\sin\left(\theta\right)\mathrm{d}\theta$. The absorption may be computed from $\sigma_{\rm abs}=\sigma_{\rm ext}-\sigma_{\rm sca}$ with:
\begin{equation}
\sigma_{\rm sca}=\frac{2\pi}{k^2}\sum_{n=1}^{\infty}\left(2n+1\right) |g_n|^2 \left(|a_n|^2+|b_n|^2\right),\,\,\,\,
\sigma_{\rm ext}=\frac{2\pi}{k^2}\sum_{n=1}^{\infty}\left(2n+1\right) |g_n|^2 \,\textnormal{Re}\,\left(a_n+b_n\right).
\label{OnAxissigmaAbs}
\end{equation}

For a Gaussian beam (on-axis) one may calculate the integrated flux of the collected beam directly, i.e. $\int\mathbf{S}^{i} \cdot \mathrm{d}\mathbf{A}$ with $2S_{\perp}^i=\textnormal{Re}(E_\theta^i H_\phi^{i*}-E_\phi^i H_\theta^{i*})/2$. The result may be written as a Cauchy-sum $\sigma_{\rm inc}^d= \frac{\pi}{2k^2}\sum_{n=1}^{\infty} \sigma_{{\rm inc},n}^d$ with summands $\sigma_{{\rm inc},n}^d$ given by:
\begin{equation}
\sigma_{{\rm inc},n}^d=\sum_{m=1}^{n}N_m g_m N_{n-m+1}g_{n-m+1}^{*}\int_{0}^{\theta_m}\left[\Sigma_m\Sigma_{n-m+1}-\left(-1\right)^{n}\Delta_m\Delta_{n-m+1}\right]\sin\theta\mathrm{d}\theta,
\end{equation}

wherein $\Delta_n\equiv \Pi_n- \tau_n$ and $\Sigma_n\equiv \Pi_n+ \tau_n$. To ensure numerical stability for small angles, a direct recursive determination of $\Delta_n\equiv \Pi_n- \tau_n$ may be used \cite{Meeten1984}. The above equations may now be used to obtain the angular spectrum of the difference signal, i.e. the PT signal analogously defined to Eqn.\ \ref{Eq:Signal}.

\begin{equation}
S\left(\theta\right)=\frac{\mathrm{d}P_d^{\rm hot} - \mathrm{d}P_d^{\rm cold}}{\mathrm{d}P_{\rm inc}^{\rm cold}}=\frac{\left[\mathrm{d}\sigma_{\rm sca}\left(\theta\right)-\mathrm{d}\sigma_{\rm ext}\left(\theta\right)\right]^{n\left(r\right),n_p}_{a_n^{L+1},b_n^{L+1}}-\left[\mathrm{d}\sigma_{\rm sca}\left(\theta\right)-\mathrm{d}\sigma_{\rm ext}\left(\theta\right)\right]^{n_0,n_p}_{a_n,b_n}}{\mathrm{d}\sigma^d_{\rm inc}\left(\theta\right)}\label{eq:GLMTPTSignal}
\end{equation}

\begin{figure}[h]
  \centering
 	\includegraphics[scale=0.75]{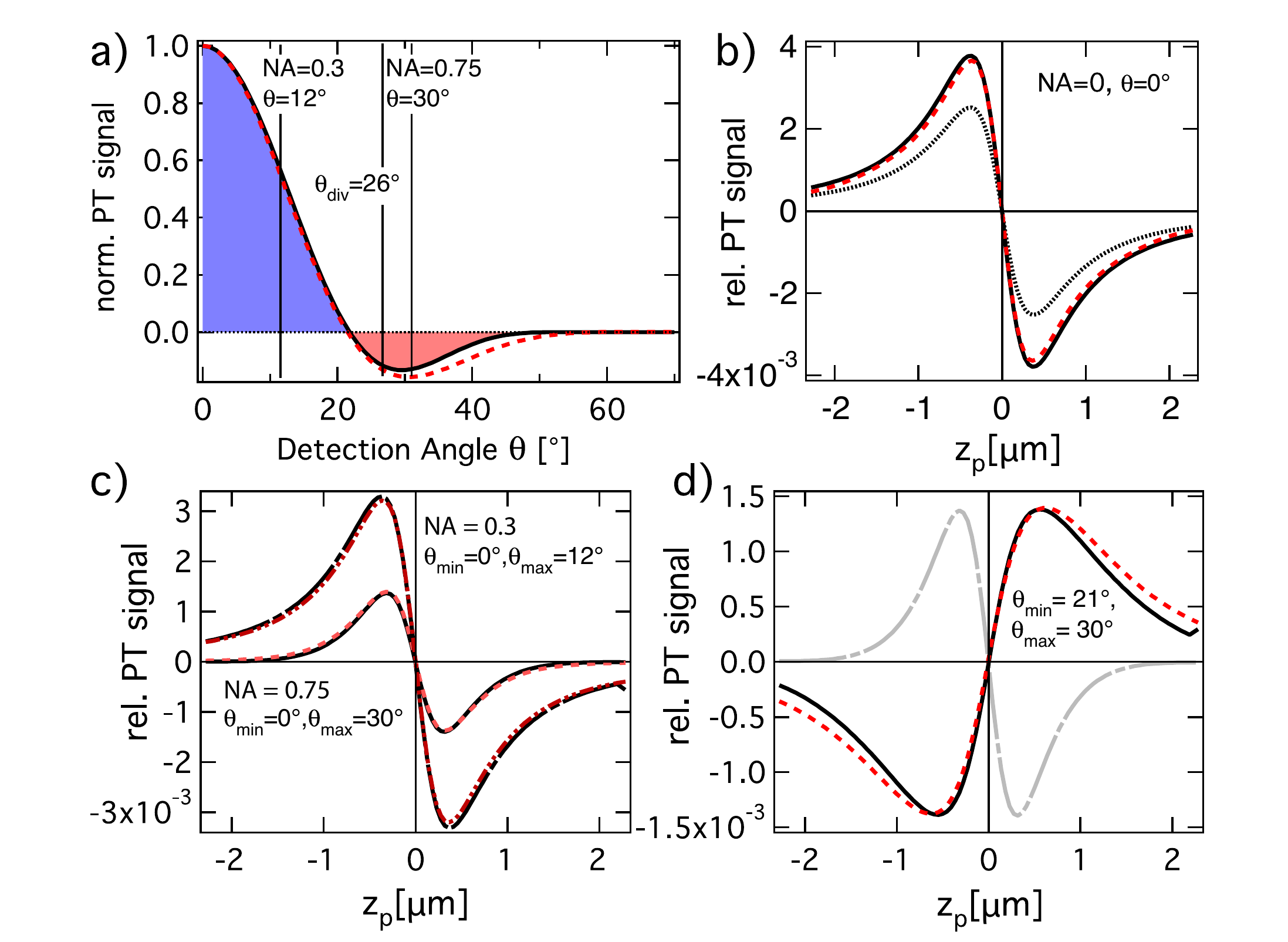}
  \caption{Comparison of the diffraction (black) and Gaussian GLMT model (red). $R=10$, $\Delta T_0=100\,\rm K$, $n_0=1.46$, $\mathrm{d}n/\mathrm{d}T=-3.6\times 10^{-3}$, $\lambda=635\,\rm nm$, $\omega_{0}=281\,\rm nm$, $\lambda_h=532\,\rm nm$, $\omega_{0,h}=233\,\rm nm$. The Diffraction model results have been scaled by a constant factor of 1.5 except for the d) where the factor is 0.8. a) PT signal angular distribution with positive (blue) and negative (red) signal. b) On-axis z-scan ($\mathrm{NA}=0, \theta=0^\circ$) of PT signal. The black dotted curve is the unscaled prediction of the diffraction model.  c) On-axis z-scan for a finite NA detection. d) On-axis z-scan with central beam stop (inverse aperture, see Figure \ref{fig:fig5Exp}). The grey curve corresponds to no central beam-stop ($\mathrm{NA}=0.75$ from c) ).  }
  \label{fig:fig3Compare}
\end{figure}

To obtain the total relative PT signal collected with a finite detection aperture, the total cross-sections $\sigma_{\rm sca, ext}$ have to be used instead of the differential ones. This corresponds then to Eqn. \ref{EqSignalNAd} in the diffraction model. The predictions of both models are displayed in Figure \ref{fig:fig3Compare}. The best agreement is found in the low-focusing regime ($\omega_0 \gg \lambda$). In the strong focusing case, the GLMT may be expected to deviate since the beam shape coefficients used here rely on the low focusing expansion of the coaxial field, while within the diffraction treatment the straight ray phase advance approximation (Eqn.\ \ref{EqPhaseAdvance}) becomes less applicable. Both frameworks agree for typical experimental parameters within a factor of the order of unity. A direct scaling-free comparison of the on-axis diffraction formula Eqn.\ \ref{EqSignalOnAxis} and the Gaussian GLMT prediction with Eqn.\ \ref{eq:GLMTPTSignal} is shown in Figure \ref{fig:fig3Compare} b), where the solid thin black line corresponds to Eqn.\ \ref{EqSignalOnAxis} and the red dashed line corresponds to Eqn.\ \ref{eq:GLMTPTSignal}. The qualitative agreement is illustrated by further z-scan examples with finite numerical collection apertures (see Figure  \ref{fig:fig3Compare} c), $\mathrm{NA}=0.3$ and $\mathrm{NA}=0.75$), i.e. Eqn.\ \ref{EqSignalNAd}, scaled by $1.5$ to match the GLMT predictions. The angular pattern in Figure \ref{fig:fig3Compare} a) of the photothermal signal has two consequences: The relative change in intensity as compared to the unperturbed beam, i.e. the relative photothermal signal, is maximal, if the detection is on-axis or in a small angular detection domain around the forward direction (see Figure\ \ref{fig:fig3Compare} b) and c) ). Further, both frameworks predict a signal inversion for the collection of an annular region of detection angles (Figure \ref{fig:fig3Compare} d)), for example for $\left[\theta_{\rm min},\theta_{\rm max}\right]=\left[21^{\circ},31^{\circ}\right]$. To test this prediction, a central beam-stop experiment has been carried out and is detailed in the following section. 
In summary, the comparison of the simple and intuitive diffraction-picture, as put forward in the preceding section, and the rigorous vectorial treatment of a Gaussian beam scattered by the GRIN lens shows a near-perfect qualitative agreement, while a quantitative agreement within a factor of order unity is found.

\clearpage

\section{Signal inversion of an axial single particle scan}
Having demonstrated the equivalence of the results obtainable within the generalized Lorenz-Mie framework and in the simple diffraction model, the following part of the paper will show a direct and obvious consequence of the above findings. Whereas the collection of the probe-beam in an angular domain around the forward-direction will show a dispersion-like lens-signature as displayed in Figure \ref{fig:fig1zScan}, the collection of an annular region can invert the detected signal. Indeed, the measurement of the photothermal signal of a single gold nanoparticle of $R=30 \, \rm nm$ radius shows this effect. To this end, a central beam-stop was introduced into the detection beam path. The collimated beam of diameter $D_o=10\,\rm mm$ was thereby reduced to an annular ring of $D_i=9\,\rm mm$ inner diameter corresponding to an angle of $\theta_{\rm min}=\arcsin\left({\rm NA}_d D_i/\left(D_o n_m\right)\right)=27^{\circ}$ as given by the Abbe sine condition and the used numerical aperture of the detection objective. The illumination objective was a high NA oil immersion objective and the resulting probe and heating beam waists were $\omega_{0,d}=281\,\rm nm$ and $\omega_{0,h}=233\,\rm nm$, respectively. The beams were offset in axial direction by $\Delta z_f=350\,\rm nm$ to achieve a symmetric signal configuration with the aberrated beams. The observed z- and xz-scans (Figure \ref{fig:fig42DxzCompare} a)) of the signal measured correspond very well to the calculated scans (Figure \ref{fig:fig42DxzCompare} b)) and clearly show the inversion of the signal. In addition to a pure change in sign of the signal, the zero-crossing shifts. Although these details are only reconcilable with the predictions of the exact GLMT description (see supplement of \cite{2011arXiv1105.3815S}), which includes the details of the aberrated beams, the basic principle and the prediction of the effect is understood based on the angular diffraction pattern described in section \ref{diffraction}.

\begin{figure}[h]
  \centering
 	\includegraphics[scale=0.2]{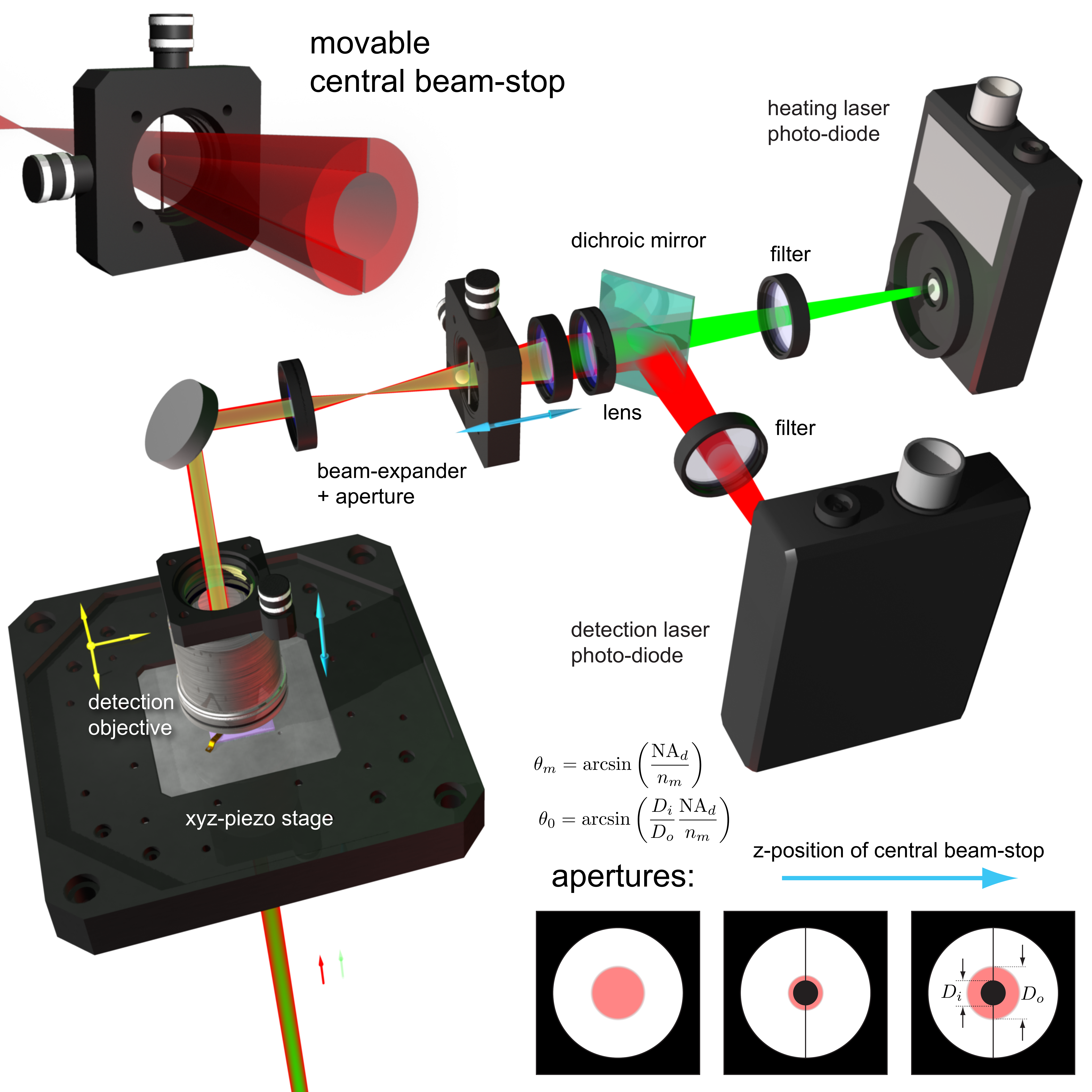}
  \caption{Experimental configuration for the central beam-stop experiment used to measure the photothermal signal displayed in Figure \ref{fig:fig42DxzCompare}. Apart from the central beam-stop,  a detailed description of the experimental setup is given in reference \cite{2011arXiv1105.3815S}.}
  \label{fig:fig5Exp}
\end{figure}

\begin{figure}[h]
  \centering
 	\includegraphics[scale=0.8]{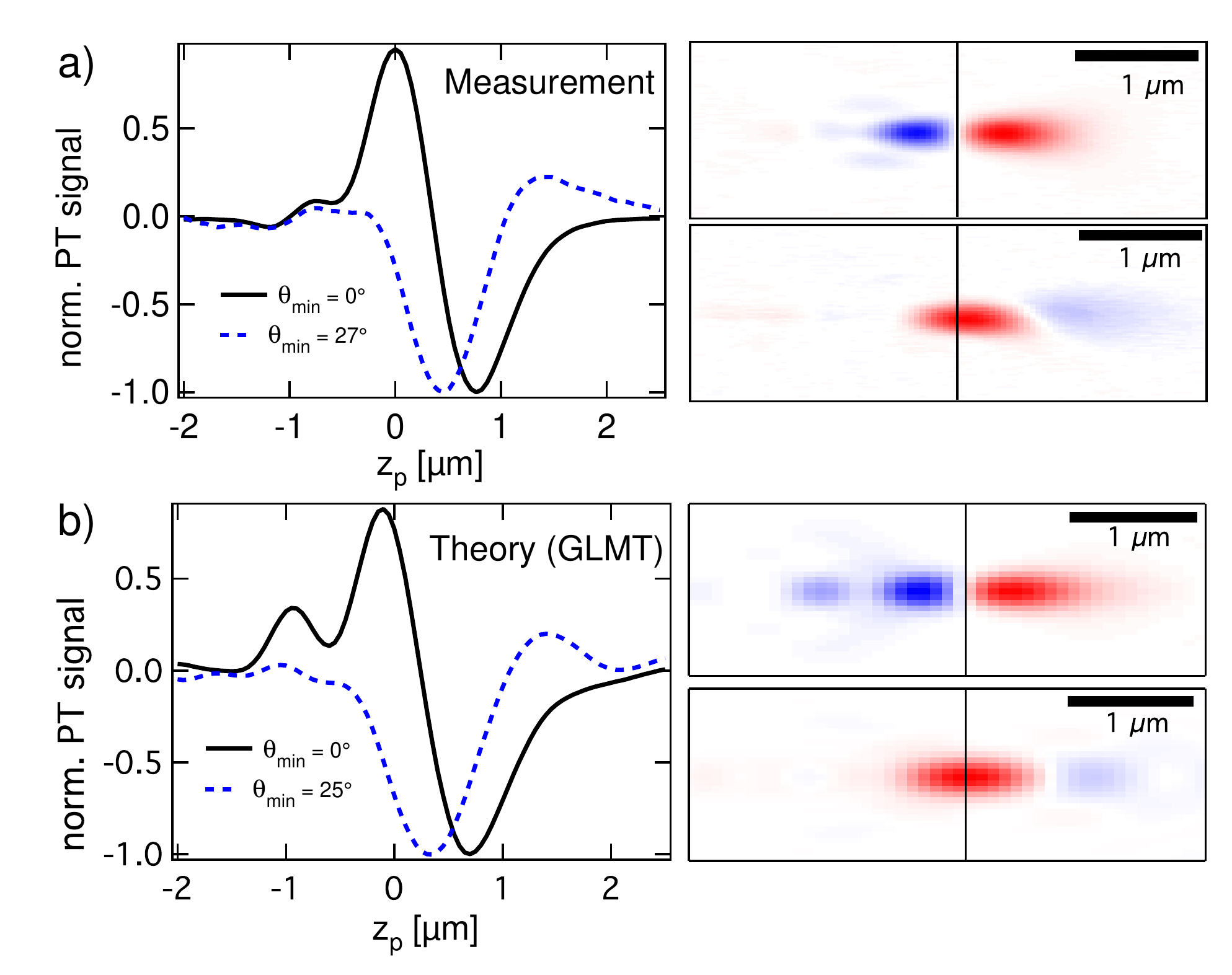}
  \caption{Photothermal Single Particle Signal z-scan (left) and xz-scan (right) measured (a) and computed (b) with the exact beam shape coefficients in the GLMT framework for a full $\mathrm{NA}=0.75$ detection (black curve on the left and upper pictures on the right) and with a inverse aperture up to an detection angle of $\theta_m$ (dashed blue on the left and lower pictures on the right). The parameters used in the calculation are the same as those in our reference \cite{2011arXiv1105.3815S}. }
  \label{fig:fig42DxzCompare}
\end{figure}

\clearpage
\section{Conclusion}
While a close relation of Gaussian beam diffraction and scattering has been shown by J. Lock et al.\ \cite{Lock1993} for the case of spherical dielectric particles, the more complicated scattering approach was thus far the only theoretical approach for the emerging technique of single particle photothermal microscopy. Although this ansatz may be used, the signature of a simple lensing mechanism (e.g. in z-scans) suggests a more intuitive model. Employing the scalar diffraction formalism common in thin sample slab absorption spectroscopy we have demonstrated that the signal obtained in photothermal single (nano-)particle microscopy can indeed be understood as the signal of a phase-modifying element, i.e. a lens, despite the microscopic origin of the latter. In contrast to the former, the spherical symmetry of the heated single-particle lens allows a direct analytical evaluation of the relative photothermal signal. The analytical model presented here could be verified in shape and absolute value when compared to experiments and a full electromagnetic vectorial treatment as published by the authors earlier. Further, it could be shown, that the Gouy-phase of a Gaussian detection beam does not contribute to the relative photothermal signal in forward direction. The understanding of the mechanism and angular distribution of the photothermal signal was then shown to explain the observed signal inversion upon the collection of an angular domain corresponding to the outer angles only. This introduces a simple and intuitive model for single particle absorption measurements and allows the quantitative assessment of nano-particle absorption cross sections.

\bibliographystyle{apsrev}

\end{document}